\documentclass{emulateapj}\usepackage{apjfonts}
\usepackage{graphicx}
\newcommand{\expnt}[2]{\ensuremath{#1 \times 10^{#2}}}   
\newcommand{\chandra}{\textit{Chandra}}
\newcommand{\xmm}{\textit{XMM}}
\newcommand{\rosat}{\textit{ROSAT}}
\newcommand{\rxj}{RX~J0720.4\ensuremath{-}3125}
\newcommand{\rbs}{RX~J1308.6+2127}

\defcitealias{kvk05}{KvK05}
\defcitealias{htdv+06}{H06}

\begin{document}

\title{Spectral and Rotational Changes in the Isolated Neutron Star
RX J0720.4$-$3125}

\author{Marten H. van Kerkwijk\altaffilmark{1},
        David L. Kaplan\altaffilmark{2},  
        George G. Pavlov\altaffilmark{3}, 
    and Kaya Mori\altaffilmark{1}}

\submitted{Accepted for publication in Astrophysical Journal (Letters), March 13, 2007}

\altaffiltext{1}{Department of Astronomy and Astrophysics, University
  of Toronto, 50 St.\ George Street, Toronto, ON M5S 3H4, Canada;
  mhvk,mori@astro.utoronto.ca.}
\altaffiltext{2}{Pappalardo Fellow; Kavli Institute for Astrophysics
  and Space Research and Department of Physics, Massachusetts
  Institute of Technology, Cambridge, MA 02139; dlk@space.mit.edu.}
\altaffiltext{3}{Department of Astronomy and Astrophysics, Pennsylvania State
  University, 525 Davey Laboratory, University Park, PA 16802;
  pavlov@astro.psu.edu}  

\begin{abstract}
\rxj\ is an isolated neutron star that, uniquely in its class, has
shown changes in its thermal X-ray spectrum.  We use new spectra taken
with \chandra's Low Energy Transmission Grating Spectrometer, as well
as archival observations, to try to understand the timescale and
nature of these changes.  We construct lightcurves, which show both
small, slow variations on a timescale of years, and a larger event
that occurred more quickly, within half a year.  From timing, we find
evidence for a `glitch' coincident with this larger event, with a
fractional increase in spin frequency of $\sim\!5\times10^{-8}$.  We
compare the `before' and `after' spectra with those from \rbs, an
isolated neutron star with similar temperature and magnetic field
strength, but with a much stronger absorption feature in its spectrum.
We find that the `after' spectrum can be represented remarkably well
by the superposition of the `before' spectrum, scaled by two thirds,
and the spectrum of \rbs, thus suggesting that the event affected
approximately one third of the surface.  We speculate the event
reflects a change in surface composition caused by, e.g., an accretion
episode.
\end{abstract}


\section{Introduction}
\label{sec:intro}

Among nearby neutron stars, seven show predominantly thermal emission,
with inferred temperatures of $kT=50$ to 100\,eV.  These so-called
isolated neutron stars (INS) are thought to be young,
$\lesssim\!1\,$Myr old, with the emission due to residual heat.
Compared to similarly aged pulsars, they have much weaker non-thermal
emission, longer, 3--10\,s periods, and stronger, few $10^{13}\,$G
magnetic fields (see \citealt{haberl06,vkk06} for recent reviews).

The INS have attracted much attention, in part because of the hope
that the equation of state in the ultra-dense interior could be
constrained using properties derived from their spectra, such as
effective temperature, surface gravity, and gravitational redshift.
Progress has been stymied, however, by difficulties in interpreting
the spectra: at present, hydrogen, helium, and mid-Z elements, in
states ranging from gaseous to condensed, all being considered (see,
e.g., contributions to \citealt*{ptz06}).

A source that may elucidate matters is \object[RX J0720.4-3125]{\rxj}.
This source was discovered with \rosat, and has a 8.39\,s period
\citep{hmb+97}.  Initial {\em XMM-Newton} and \chandra\ spectra
appeared to be featureless, blackbody-like \citep{pmm+01,kvkm+03},
similar to those of the brightest INS, \object[RX J1856.5-3754]{RX
J1856.5$-$3754} \citep[e.g.,][]{bzn+01}, but more detailed analysis
uncovered a weak, broad absorption feature near 280\,eV
\citep{hztb04}.  Similarly broad (but stronger) features at energies
from 200 to 700\,eV were found in five other INS
\citep{hsh+03,hmz+04,vkkd+04,zct+05} and are generally interpreted as
proton cyclotron or neutral-hydrogen transitions.  For \rxj, either
interpretation is consistent with the magnetic field strength
$B\simeq2\times10^{13}\,$G inferred from timing \citep[ hereafter
\citetalias{kvk05}]{kvk05}.

Unlike other INS, however, the spectrum of \rxj\ changed
\citep{dvvmv04}.  The change could be described as an increase in
temperature, from 86 to $94\,$eV, and an increase in equivalent width of the
absorption feature, from $\sim\!5$ to $70\,$eV \citep[][ hereafter
\citetalias{htdv+06}]{htdv+06}.  The nature of the changes is unclear.
\citet{dvvmv04} and \citet{vdvmv04} find that they started gradually
and then speeded up, and \citetalias{htdv+06} find a recent, partial
reversal.  These authors interpret this as evidence for precession on
a $\sim\!7\,$yr period.

In this {\em Letter}, we reconsider the timescale of the changes,
investigate what changed, and offer an alternative interpretation.
After describing the data we use in \S\,\ref{sec:obs}, we present
lightcurves in \S\,\ref{sec:lightcurves} and update the timing
solution of \citetalias{kvk05} in \S\,\ref{sec:timing}, finding
evidence for a glitch at the time of the largest spectral change.  We
study the `before' and `after' spectra in \S\,\ref{sec:spectra}, and
suggest that only part of the surface changed.  We discuss
implications in \S\,\ref{sec:disc}.

\section{Observations, Reduction, and Analysis}
\label{sec:obs}

We used \chandra\ to observe \rxj\ several times in 2005 and 2006, for
$\sim\!300\,$ks total, taking spectra with the Low Energy Transmission
Grating Spectrometer (LETG).
For comparison, we analysed archival spectra taken in 2000
($\sim\!38\,$ks)
and 2004 ($\sim\!36\,$ks),
as well as spectra of another INS, \rbs, taken in
\dataset[ADS/Sa.CXO#4595]{2004} ($\sim\!90\,$ks).

For all observations, we extracted spectra following standard
procedures (we used CIAO ver.\ 3.3 and CALDB ver.\ 3.2).  We
calibrated the spectra outside of CIAO: We subtracted the background,
corrected for higher-order contributions using the observed counts and
the Nov.\ 2004 effective
areas,\footnote{\url{http://cxc.harvard.edu/cal/Letg/Hrc\_QE/ea\_index.html}}
and calculated photon rates using the first-order effective areas.
Note that we implicitly assume diagonal response matrices, thus
ignoring the LETG line-spread function; this should be fine for
spectra without sharp features, like ours.  Finally, for timing
purposes, we extracted barycentered source events (following
\citetalias{kvk05}).

We augmented our data with archival \xmm\ observations.  For timing
purposes, we extracted barycentered source events from all EPIC
instruments (following \citetalias{kvk05}), while for making
lightcurves we extracted RGS spectra.  The latter was done with {\tt
rgsproc} (XMMSAS ver.\ 7.0), which was also used to produce
flux-calibrated spectra (ignoring, like for LETG, the line-spread
function).  We note that XMMSAS ver.\ 7.0 includes corrections for the
changing long-wavelength sensitivity of RGS,\footnote{See XMM-SAS
ver.\ 7.0 and XMM-CCF-REL-216 release notes.} which has influenced
previous analyses (see \S~\ref{sec:lightcurves}).

\begin{figure}
\includegraphics[width=\hsize]{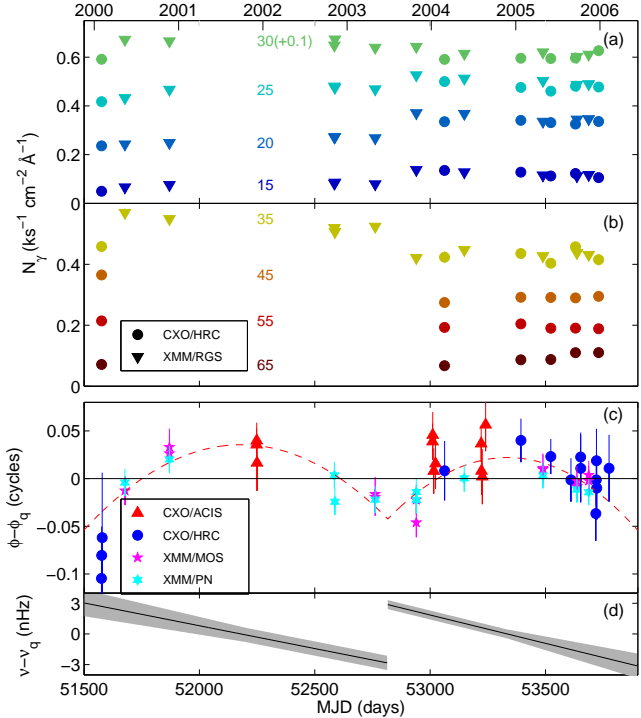}
\caption{Lightcurves and timing residuals for \rxj.  Panels {\bf a}
  and {\bf b}: Photon rates in various wavelength bands.  The central
  wavelengths are indicated; the widths are 5\,\AA\ for bands between
  15 and 35\,\AA, and 10\,\AA\ for those from 45 to 65\,\AA\ (the
  latter bands are not covered by RGS).  For clarity, the 30\,\AA\
  fluxes are offset vertically by 0.1 unit.  The formal uncertainties
  are smaller than the symbol size.  Panel {\bf c}: Arrival-time
  residuals relative to a simple spin-down model (fit to \chandra\ and
  \xmm\ data only).  The different instruments are indicated, and the
  uncertainties include, in quadrature, a systematic term of 0.11\,s.
  Overdrawn (dashed red curve) is the best-fit glitch model, which
  reproduces the data much better.  Panel {\bf d}: Difference in spin
  frequency of the glitch model relative to the simple spin-down model
  (solid curve with 1$\sigma$ uncertainties indicated in grey).
\label{fig:lightcurves}
\label{fig:timing}
}
\end{figure}

\section{Time Scales of the Spectral Change}
\label{sec:lightcurves}

To determine the timescale on which the spectral changes occurred, we
used the LETG and RGS spectra to produce lightcurves in eight
wavelength bands (Fig.~\ref{fig:lightcurves}).  Broadly, these confirm
previous results (see \S\,\ref{sec:intro}): as time passed, the flux
increased at wavelengths $\lambda\lesssim25\,$\AA, and decreased in
the range $35\lesssim\lambda\lesssim55\,$\AA, presumably because the
absorption feature deepened.  However, with the new RGS
long-wavelength efficiencies, the evidence for a decrease in
25--36\,\AA\ flux before 2003 has largely disappeared (but the first
LETG and RGS observations have become discrepant, suggesting some
calibration issues remain).

Figure \ref{fig:lightcurves} shows that most of the change occurred
rapidly, in between the 2003 May and 2003 October \xmm\ observations.
Before and after, there are slower variations, with those before in
the same sense as the rapid change, and those after in the opposite
one.  This is consistent with inferences from PN data by \citetalias{htdv+06}.

To examine broad-band changes, we integrated the LETG fluxes in the
10--80\,\AA\ range (0.155--1.25\,keV), finding totals of 0.97, 1.06,
1.04, 1.02, 1.04, and $1.03\times10^{-11}{\rm\,erg\,s^{-1}\,cm^{-2}}$
for the six observations, respectively (with $\sim\!2\%$ uncertainty).
Thus, from 2000 to 2004 the flux increased by $\sim\!9\%$, and
afterwards it possibly decreased slightly.  The change of the total
source flux is likely similar: most flux is emitted in the
10--80\,\AA\ band, and, despite the change in spectral shape,
different epochs have extinction corrections identical to
$\lesssim\,1\%$ (e.g., factor 1.43 for $N_{\rm
H}=1.3\times10^{20}{\rm\,cm^{-2}}$).

\citetalias{htdv+06} suggested the changes were cyclic, with a
$\sim\!7\,$yr period, but the lightcurves do not require this.
Rather, to us Fig.~\ref{fig:lightcurves} (as well as Fig.~2 of
\citetalias{htdv+06}) suggests a major, impulsive change in 2003, the
source `building up' to the change beforehand, and `relaxing back'
afterwards.  

\begin{deluxetable}{l@{~}r@{~~}l@{~}r@{~~}r@{~~~}l}
\tablecaption{Log of New Observations and Times of Arrival\label{tab:obs}}
\tablehead{
&&&\colhead{Exp.}&& \colhead{TOA\tablenotemark{c}}\\
\colhead{Instrument\tablenotemark{a}}&
\colhead{ID\tablenotemark{b}}&
\colhead{Date}&
\colhead{(ks)}&
\colhead{Counts}&
\colhead{(MJD)}
}
\startdata
PN\dotfill &   \dataset[ADS/Sa.XMM#x/0300520201]{986-S3} &  2005 Apr 28 & 52.0 & 300204 & 53488.6756038(3)\\
MOS1\dotfill & \dataset[ADS/Sa.XMM#x/0300520201]{986-S1} &              & 41.1 & 51871  & 53488.6013068(8)\\
MOS2\dotfill & \dataset[ADS/Sa.XMM#x/0300520201]{986-S2} &              & 41.2 & 54046  & 53488.6012096(8)\\
HRC\dotfill &  \dataset[ADS/Sa.CXO#5582]{5582} &            2005 Jun 01 & 69.7 & 903822 & 53522.9397230(12)\\
            &  \dataset[ADS/Sa.CXO#6364]{6364} &            2005 Aug 27 & 38.6 & 520744 & 53610.0880958(18)\\
PN\dotfill &   \dataset[ADS/Sa.XMM#x/0300520301]{1060-S3} & 2005 Sep 23 & 51.0 & 284343 & 53636.3001458(4)\\
MOS1\dotfill & \dataset[ADS/Sa.XMM#x/0300520301]{1060-S1} &             & 36.9 & 50393  & 53636.2024428(8)\\
MOS2\dotfill & \dataset[ADS/Sa.XMM#x/0300520301]{1060-S2} &             & 36.9 & 51231  & 53636.2027347(8)\\
HRC\dotfill &  \dataset[ADS/Sa.CXO#6369]{6369} &            2005 Oct 08 & 26.0 & 409201 & 53652.0859352(21)\\
            &  \dataset[ADS/Sa.CXO#7177]{7177} &            2005 Oct 09 & 8.0  & 106581 & 53652.5397755(32)\\
PN\dotfill &   \dataset[ADS/Sa.XMM#x/0311590101]{1086-S3} & 2005 Nov 12 & 38.0 & 268168 & 53687.1718885(4)\\
MOS1\dotfill & \dataset[ADS/Sa.XMM#x/0311590101]{1086-S1} &             & 39.5 & 56034  & 53687.1658660(8)\\
MOS2\dotfill & \dataset[ADS/Sa.XMM#x/0311590101]{1086-S2} &             & 39.5 & 56292  & 53687.1658654(7)\\
HRC\dotfill &  \dataset[ADS/Sa.CXO#7243]{7243} &            2005 Dec 14 & 17.1 & 241644 & 53718.1462899(25)\\
            &  \dataset[ADS/Sa.CXO#7244]{7244} &            2005 Dec 15 & 16.2 & 227541 & 53719.8089713(22)\\
            &  \dataset[ADS/Sa.CXO#7245]{7245} &            2005 Dec 16 & 17.1 & 249521 & 53720.7701604(30)\\
            &  \dataset[ADS/Sa.CXO#5584]{5584} &            2005 Dec 17 & 14.1 & 194582 & 53721.9195718(30)\\
            &\dataset[ADS/Sa.CXO#7251]{7251} &              2006 Feb 09 & 10.6 & 160780 & 53775.3508981(32)
\enddata
\tablecomments{A log of earlier observations and corresponding times
  of arrival can be found in \citet{kvk05}.}
\tablenotetext{a}{
  HRC: High-Resolution Camera for spectroscopy aboard \chandra,
       used with the Low-Energy Transmission Grating.
  PN: European Photon Imaging Camera (EPIC) with PN detectors aboard \xmm,  
      used in full-frame mode with a thin filter.
  MOS1/2: EPIC with MOS detectors aboard \xmm,
      used in small-window mode with thin filter.}
\tablenotetext{b}{Observation identifier (\chandra) or revolution
  number and exposure identifier (\xmm).}
\tablenotetext{c}{Time of arrival of maximum light closest to the
  middle of an observation, as determined from fitting a sinusoid to
  the binned lightcurve; it is given with 1-$\sigma$ uncertainties.}
\end{deluxetable}

\section{A Change in Timing Properties}
\label{sec:timing}
We updated the timing solution of \citetalias{kvk05}, determining
arrival times for all new observations (see Table~\ref{tab:obs}).
These were consistent with the old timing solution, but, as before,
significant timing residuals remained (Fig.~\ref{fig:timing}): 
to obtain $\chi^2_{\rm red}=1$
required adding (in quadrature) a systematic error term of
$\sim\!0.3$~s.

Inspired by what to us seems a sudden spectral change
(\S~\ref{sec:lightcurves}), we considered a `glitch' as the source of
the timing residuals.  Earlier, we had found some improvement with a
rudimentary glitch model, in which the frequency $\nu$ changed after
mid-2003 and recovered only slightly \citepalias{kvk05}.  Here, we
tried a different model, in which we fit for different $\nu$ and
$\dot\nu$ before and after a given glitch time $t_g$, but keep phase
continuous, ignore any recovery, and leave the cycle counts unchanged
from the simple spin-down model.  This improves the fit drastically
(Fig.~\ref{fig:timing}): to get $\chi^2_{\rm red}\simeq1$, a
systematic error term of only 0.11~s is required.  Intriguingly, the
best-fit glitch time coincides with the period of the greatest
spectral change.

In Table~\ref{tab:timing}, we list parameters for the simple spin-down
and the glitch models.  We try solutions both with and without the
\rosat\ data (which have much less secure cycle counts); in
either case, the change in $\nu$ is significant, but that in $\dot\nu$
is not.

\begin{deluxetable}{l c c}
\tablecaption{Timing Solutions for \rxj\label{tab:timing}}
\tablewidth{0pt}
\tablehead{
\colhead{Quantity} & \colhead{Excl.\ \rosat} & \colhead{All Data}
}
\startdata
\sidehead{Spindown only}
$t_0$ (MJD) &53010.2635646(7) &53010.2635626(6)\\
$\nu$ (Hz) &0.11917366979(12) &0.11917366954(11)\\
$\dot \nu$ ($\mbox{Hz s}^{-1}$) &$\expnt{-9.74(4)}{-16}$ & $\expnt{-9.88(13)}{-16}$\\
TOA rms (s) & 0.26 & 0.29\\
$\chi^2/\mbox{dof}$ & 77.6/46=1.69& 150.8/49=3.08\\
\sidehead{Spin-down + Glitch}
$t_{0}$ (MJD) &53010.2635686(10) & 53010.2635667(10)\\
$\nu$ (Hz) & 0.1191736716(9) & 0.1191736716(9) \\
$\dot \nu$ ($10^{-15}\,\mbox{Hz s}^{-1}$) &$-1.04(3)$ &$-1.04(3)$\\
$t_g$ (MJD) &52817(61) &52866(73)\\
$\Delta\nu$ (nHz) & 5.7(17) & 4.1(12)\\
$\Delta\dot\nu$ ($10^{-17}\mbox{ Hz s}^{-1}$) & $-1(4)$ & $-4(3)$\\
TOA rms (s) & 0.15 &0.24\\
$\chi^2/\mbox{dof}$ & 37.0/43=0.86&$45.1/46=0.98$
\enddata
\tablecomments{The parameters determine the cycle count plus phase via
  $\phi(t)=\nu(t-t_0)+\frac{1}{2}\dot\nu(t-t_0)^2+\Delta\phi_g(t)$,
  where
  $\Delta\phi_g(t)=-\Delta\nu(t-t_g)-\frac{1}{2}\Delta\dot\nu(t-t_g)^2$
  for $t<t_g$ in the glitch model and zero otherwise.  For all fits, a
  0.11~s systematic uncertainty has been added in quadrature to the
  times of arrival (TOAs), and the uncertainties quoted are twice the
  formal 1$\sigma$ values.}
\end{deluxetable}

The addition of a glitch improves the fit, but also adds complexity.
To evaluate the significance of the improvement, we tried two other
ways of adding complexity.  First, we included higher-order frequency
derivatives, but found a similarly improved fit required eighth order
(i.e., six additional parameters).  Second, motivated by the
suggestion of \citetalias{htdv+06} of a $\sim\!7\,$yr periodicity in
the spectral variations, we added a periodic component, but found a
similarly improved fit only for periods near 4\,yr (as expected from
Fig.~\ref{fig:timing}; for details, see \citealt{vkk06}).  The result
did not seem to depend on whether or not we included the \rosat\ data
or changed their cycle counts (as suggested by \citealt{haberl06}).
Overall, the timing seems to support an impulsive model, but does not
exclude periodic changes.

\begin{figure}
\includegraphics[width=\hsize]{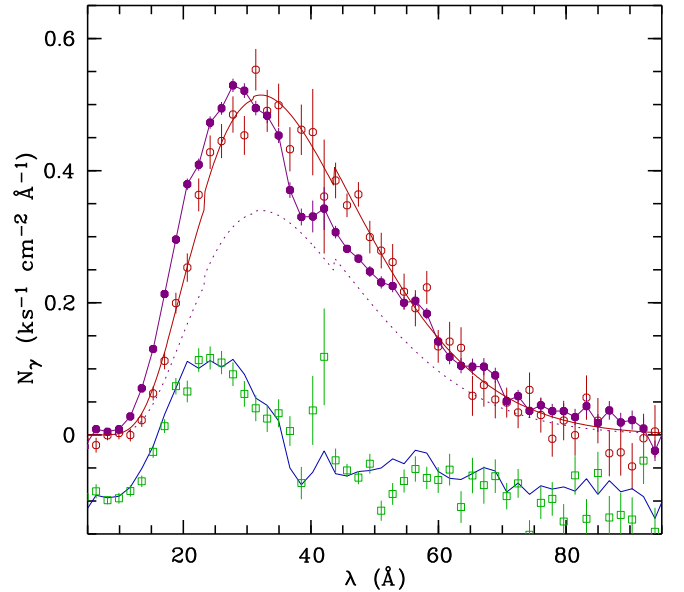}
\caption{Calibrated LETG spectra.  The spectrum of \rxj\ from 2000
   (red open circles) is well described by an absorbed black-body
   ($N_{\rm H}=(1.55\pm0.11)\times10^{20}{\rm\,cm^{-2}}$,
   $kT=83.3\pm1.1$\,eV, and $r/d=0.0189\pm0.0008{\rm\,km\,pc^{-1}}$;
   drawn red curve).  The average post-2003 spectrum (magenta filled
   circles) could be a superposition of spectra from unchanged and
   changed parts of the surface.  As an indication, a close to maximum
   contribution by the unchanged spectrum is indicated (magenta,
   dotted line).  The residual (blue, wiggly curve, offset down by 0.1
   unit for clarity) is remarkably similar to the spectrum of \rbs\
   (green squares, scaled by a factor 1.2).  For all spectra, the
   40--44\,\AA\ range is less reliable, because of strong instrumental
   carbon absorption.
\label{fig:spectra}
}
\end{figure}

\section{Nature of the Spectral Change}
\label{sec:spectra}

The lightcurves show that most of the spectral change in \rxj\ took place
in 2003.  In Fig.~\ref{fig:spectra}, we show
the LETG spectrum from 2000, taken before this time, as well
as the average of all LETG spectra taken after 2003 (the further small
changes in 2004 and 2005 [Fig.~\ref{fig:lightcurves}] would not be
noticeable in Fig.~\ref{fig:spectra}).  For comparison, we also show
the LETG spectrum of \rbs, which is very similar to \rxj\ in its
inferred temperature, extinction, and magnetic field strength
($kT\simeq100\,$eV, $N_{\rm H}\simeq 2\times10^{20}{\rm\,cm^{-2}}$,
\citealt{shhm06}; $B\simeq3\times10^{13}\,$G, \citealt{kvk05b}).

The comparison of the spectra raises two main questions.  First, for
\rxj, how can the appearance of a pronounced absorption feature be
associated with only a small change in temperature (from $kT\simeq86$
to $94~$eV; \citetalias{htdv+06})? Second, why is the absorption
feature in \rbs\ much stronger than that in \rxj\ (in either state)?

Of these two questions, we believe the answer to the second may be
simple: that not the whole but only part of the surface changed.  If
so, then the `after' spectrum contains a contribution from unchanged
parts that emit the cooler, nearly featureless `before' spectrum.  Hence, the
spectrum from the changed part should be hotter and should have a
stronger line than one would infer from the average.

We do not know the relative contribution of the `before' spectrum, but
an upper limit of $\sim\!70$\% is set by the requirement that it does
not exceed the `after' spectrum at any wavelength (set by the 35--40\,\AA\ region;
Fig.~\ref{fig:spectra}).  For a contribution near this limit, the
absorption feature in the changed parts of the surface would be very
strong, just as observed in \rbs.  Indeed, if we subtract the `before'
spectrum, scaled by two thirds, from the `after' spectrum, the
residual is remarkably similar to the spectrum of \rbs\ (see
Fig.~\ref{fig:spectra}).  This is confirmed by the EPIC-pn data (Mori
et al., in preparation).

\section{Ramifications}
\label{sec:disc}

We have found that most of the spectral variations in \rxj\ occurred on a
short, less than half-a-year timescale, and that, coincident with this, 
there likely was an increase in spin frequency.  
We also showed that the `after' spectrum could be reproduced if
one third of the emission changed from the blackbody-like `before'
spectrum to something similar to the emission of \rbs,
and we suggested only part of the visible surface changed.  
Here, we will interpret these findings assuming
an impulsive event (see \citetalias{htdv+06} for a
discussion in terms of 
precession).  As the observations show evidence for a reversal, we
need to address not only the nature and cause of the change, but also
how it can be undone.

It is difficult to understand the change without knowing the
composition and state at the photosphere (\S\,\ref{sec:intro}).  The
order-of-magnitude increase in strength of the absorption feature
suggests, however, that a new species became dominant, due to a change
in ionisation balance, state (e.g., from condensed to gaseous), or
composition.

In principle, a temperature change could affect the ionisation
balance or state, and one could envisage heating due to a sudden
energy release associated with, e.g., a glitch, and subsequent
cooling on a timescale set by the release depth.  However, the
average observed flux before and after the event differs by only
$\sim\!5$\%.  Even if due to only one third of the surface, the
implied increase in effective temperature is at most~4\%
(\citetalias{htdv+06} find a 6\% change in average temperature from
spectral fits, but this may simply reflect that a model of a blackbody
with a gaussian absorption line is not realistic.)  With such a small
change, it is not clear the atmosphere and spectrum could change so
drastically.  

An easier explanation for the strengthening of the absorption feature
would be a change in composition, e.g., due to an accretion
episode.  Such a change could also lead to a small increase in
effective temperature, since heat conduction is faster for lighter
elements \citep[e.g.,][]{pycg03}.  But how to undo the change?  For
hydrogen, there is a possible mechanism \citep*{cab04}: for
$B\simeq2\times10^{13}\,$G and $T_{\rm eff}\simeq10^6\,$K, hydrogen
will be burnt diffusively on carbon in $\sim\!200\,y_{\rm
H,10}^{-5/12}\,$yr (where $y_{\rm H}=10y_{\rm H,10}{\rm\,g\,cm^{-2}}$
is the hydrogen column density), with substantial uncertainty due to
the composition of the underlying layers.  A thin hydrogen layer,
superposed on a 
condensed surface, has been invoked by \citet*{mzh03} and
\citet{hkc+06} to explain the featureless spectra of some INS.  For
\rxj, it would imply a very thin hydrogen layer before the change,
and a thicker one on parts of the surface afterwards.

Among the two possibilities mentioned above -- a glitch
in the neutron-star crust and an accretion episode -- the former would most
naturally explain the timing results; indeed, the fractional increase
in spin frequency of $\Delta\nu/\nu=5\times10^{-8}$ would not be
unusual for an older radio pulsar \citep[e.g.,][]{js06}.  Also for
magnetars, glitches are observed, as well as slow changes on
timescales of years (see \citealt{kaspi06} for a review).  The
glitch could release some energy, causing a temperature increase that
depends on the release depth (e.g.,
\citealt{hc04} and references therein).  In addition, it could induce
magnetospheric currents, which could produce hydrogen and helium
on the surface by spallation (as suggested for RX J1856.5$-$3754 by
\citealt{hkc+06}).

Accretion would naturally lead to a change in composition, as
well as, by its associated angular momentum, a change in spin.  Matter
could originate in an asteroid or comet belt such as has been invoked
to explain timing noise, nulling and other pulsar properties by
\citet{cs06}.  If accretion started due to a collision between
asteroids, low-rate accretion of smaller debris could perhaps cause
the slower changes.  The
observed frequency change requires an accreted mass of
$\sim\!10^{20}$--$10^{21}\,$g, depending on accretion regime and
properties of accreting matter (large bodies accreting directly onto
the star, small grains destructed and ionized before reaching
the Alfven surface, etc.).  The accretion would result in an energy
of $\sim\!10^{40}$--$10^{41}\,$erg being released in a relatively short time,
hence a high luminosity.  We have checked the {\em RXTE} All-Sky
Monitor (ASM) data in 2003 May-October and found no emission from
\rxj; however, given the low ASM duty cycle,
a short-lived ($\lesssim\!10\,$ks) event
could easily be missed.

In future, the above ideas -- as well as the alternative suggestion of
precession \citepalias{htdv+06} -- can be tested.  For instance, both
give different predictions for future timing.  Generally, if hydrogen
appeared on the surface (or hydrogen-covered parts became visible),
the optical flux should perhaps have increased, yet it changed by
$\lesssim\!10\%$ \citep*{kvka07}.  Also, if the composition changed
locally, e.g., near a magnetic pole, matter might diffuse
outwards, perhaps leading to further slow changes.  If so, one would
also expect that the pulse 
profile evolved, perhaps being most sharply peaked, especially at high
energies, shortly after the event.  From the PN data, this indeed
appears to be the case (Mori et al., in preparation).

\vspace*{-3mm}
\acknowledgements We thank Ron Remillard for help with the {\sl RXTE}
ASM data analysis.  Support for this work was provided by NASA
(Chandra award GO5-6074).

{\it Facilities:} \facility{CXO (LETG/HRC-S)}, 
                  \facility{XMM (RGS, EPIC)}

\end{document}